# Contrasting Analog and Digital Resistive Switching Memory Characteristics in Solution-Processed Copper (I) Thiocyanate and Its Polymer Electrolyte Based Memristive Devices

*Rajesh Deb[1], Saumya R. Mohapatra\*[1], Manjula G. Nair[2], and Ujjal Das[3]*

[1]Solid State Ionics Laboratory, Department of Physics, National Institute of Technology Silchar, Silchar-788010, Assam, India

[2]Department of Physics, Indian Institute of Technology, Patna, Bihar, India, 801106

[3]Quantum Materials and Device Unit, Institute of Nano Science and Technology, Mohali-140306, Punjab, India

*Email:saumya@phy.nits.ac.in

ABSTRACT

Usually, resistive switching (RS) devices show digital RS memory (sharp SET and RESET process), which is most suitable for digital data storage applications. Some RS devices also manifest ideal memristive behavior or analog memory characteristics (gradual change in resistance states). The analog RS properties of memristive devices widen their application domain to a much broader field of neuromorphic computing. The tunability of memristive devices to digital or analog memory applications greatly depends upon the switching medium. In this work, we report a comparative study on RS properties of two kinds of memristive devices based upon copper (I) thiocyanate (CuSCN) and a solid polymer electrolyte (SPE) made up of CuSCN as ionic moieties in polyethylene oxide (PEO). The device (ITO/CuSCN/Cu), prepared by spin-coating CuSCN layer between ITO and copper electrode, shows simultaneous analog and digital RS characteristics. The RS property of the device is tunable by varying the thickness of the CuSCN layer. The current-voltage characteristics reveal that devices prepared at 3000 rpm (thicker) during the spin-coating show only digital bipolar RS memory. In comparison, the devices deposited at 4000 rpm (thinner) show both analog and digital RS memory. The conduction mechanism responsible for RS behavior in CuSCN-based devices is Schottky emission mediated charge trapping and de-trapping at the interfacial states. Contrastingly, when the same CuSCN is used as the electrolyte in SPE film, the device only shows bipolar digital non-volatile memory characteristics. The RS behavior is due to the electrochemical metallization (ECM) mechanism. The ON and OFF states are achieved by the formation and rupture of copper filaments due to the redox reactions at the interface.

1. INTRODUCTION

With the advent and adaptation of new technologies such as the internet of things (IoT), artificial intelligence (AI), and neuromorphic computing, etc., in many walks of our life, society is increasingly becoming more data-driven. To cope with the surge of demand in storing vast amounts of data and processing it faster, the storage media, i.e., digital memory technology, is facing the challenge of improving memory density and switching speed. So far, memory density has increased in tandem with Moore's law. Hence, the cell size has been continuously downscaled and may already hit the physical limit (i.e., $4F^2$ design protocol where F is the minimum feature size in lithography).[1-3] Similarly, the slow switching speed of the memory cell creates a bottleneck between the memory and the processor in the von-Neumann architecture based platforms, where they are segregated as different units.[3-5] Hence, the overhauling of digital memory technology is eminent and rightly happening, as evident from the recent developments of in-memory-computing or near-data-processing device architectures.[5-8] In this context, the 'resistive switching memory' (RSM) brings a new prospective with versatile memory characteristics. The RSM devices, also known as memristive devices, show voltage modulated resistance states, and their memory characteristics can be broadly classified as digital and analog types. Digital RSM devices have features of sharp SET and RESET processes that can be observed in the bipolar or unipolar mode of operations. The early studies on RSM are primarily focused on observing digital non-volatile memory operations, which is useful in data storage. On the other hand, the analog RSM shows a gradual change of resistance states as the voltage is scanned. It represents ideal memristive behavior i.e., showing pinched hysteresis in the current-voltage characteristics.

Recent reports of RS devices show digital RS memory characteristics can be repurposed to observe analog RS memory, and vice-versa, simply by maneuvering the biasing conditions.[9-11]

Such versatility in the memory characteristics of memristive devices widens their application potential from conventional digital storage to the implementation of neuromorphic computing hardware and mimicking biological synaptic memory. Further, memristive devices also see many specialized niche applications with novel device architecture and by on-boarding them with other micro/nanodevices. Lately, memristor-based sensors (memsensors), where along with the electrical bias, other external stimuli also strongly influence the resistance state of the devices, are explored for various sensing applications.[12-13] Similarly, the self-powered memristive device is a new trend where hybrid device architectures are designed with nanogenerators, thin film solar cells, etc.[14-17]

Memristive devices endowed with such varied functionalities and multiple memory characteristics owe much to the switching medium. This work focuses on copper thiocyanate (CuSCN) as a promising switching medium. CuSCN is already reasonably known as an excellent electronic and optoelectronic material having applications in a wide range of devices such as perovskite photovoltaics, organic LED, deep-UV photodetectors, thin film transistors, and radiofrequency Schottky diode, etc.[18-22] Attributes such as chemical stability, optical transparency (98% in the visible range), good hole mobility (0.01–0.1 cm$^2$ V$^{-1}$ s$^{-1}$) and suitable energy band alignments to support hole transport are some of the reasons make it very popular as a hole transport layer (HTL) in these devices.[23] CuSCN is a P-type semiconductor from the family of pseudohalides and, due to the interbonding of layers, can form a coordination polymer. Moreover, CuSCN is solution processable with solvents such as diethyl sulfide (DES) and NH$_4$OH, producing large-area thin films without cracks and pin-holes.[24] Hence, it has a huge scope for low-cost solution-processable flexible and transient electronic applications. But, despite these advantages, CuSCN as the primary or sole switching material in memristive devices is less explored. Initial studies on CuSCN-based RS memristive devices used a composite of CuSCN or copper-dopped CuSCN solid electrolytes showing digital RS characteristics.[25-26] In both cases, the deposition method of CuSCN was rigorous and complicated, involving either thermal evaporation of KSCN on the copper surface or dipping in the copper electrode in a solution of NaSCN. Another work involving CuSCN was reported by B. Cheng *et al.*[27] They showed bipolar RS behavior in a multilayer RS medium made up of CuSCN/PMMA/ZnO that worked as a p-i-n heterojunction diode. More recently, W. Chen *et al.* reported negative differential resistance and bipolar resistive switching memory in symmetric ITO/CuSCN/ITO devices where the CuSCN layer was electrodeposited.[28] Here, in this study, we fabricated two types of memristive devices with the switching medium as (i). CuSCN and (ii). CuSCN-based solid polymer electrolyte (Cu-SPE). Both the switching materials are solution-processed and spin-coated. The Cu-SPE is comprised of CuSCN dissolved in polyethylene oxide (PEO). The CuSCN and Cu-SPE films are pretty different regarding their electrical properties. The CuSCN layer, though often considered a solid electrolyte, its ionic moieties are not mobile. It is purely an electronic conductor through holes.[29] But in Cu-SPE, CuSCN remains dissolved with separate and mobile ionic species ($Cu^+$ and $SCN^-$). Hence, it remains predominantly an ionic conductor with a high ion-transport number.[32] Our first device with ITO/CuSCN/Cu stacking shows both digital and analog-type resistive switching properties with the ability of synaptic plasticity. In contrast, the devices with ITO/Cu-SPE/Cu structure show only digital RS memory based on the electrochemical metallization (ECM) mechanism.

## 2. EXPERIMENTAL SECTION

### 2.1 Fabrication of CuSCN based memristive cells

Copper (I) thiocyanate (CuSCN) and Polyethylene Terephthalate (PET) substrates with Indium-tin Oxide (ITO) layer pre-deposited were brought from Sigma Aldrich. The substrate was cut into pieces of 18 ×18 mm$^2$ areas. The solution of CuSCN was prepared using diethyl sulfide as the solvent and spin-coated onto the ITO layer of the PET substrate. Devices were prepared with three different thicknesses of CuSCN layer by spin-coating at the speed of 3000, 4000, and 5000 rpm for 60 secs. The prepared films were then vacuum-dried at 60 $^0$C for 5 hours to remove the solvent. Finally, a 40 nm thick circular copper (Cu) top electrode of diameter (100 µm) was deposited using a stainless steel shadow mask with a thermal evaporation method. So, a vertical two-terminal ITO/CuSCN/Cu device was formed, where ITO and Cu act as bottom and top electrodes, respectively.

**2.2 Preparation of PEO-CuSCN Solid Polymer Electrolyte (SPE)**

Poly(ethylene oxide) (PEO) is known to solubilize many alkali and alkaline earth metals salts due to the large dipole moment on the ether oxygen.[30] Copper (I) thiocyanate as a pseudohalide is also expected to dissolve in PEO. Following the earlier reports, we prepared solid polymer electrolytes (SPEs) using PEO and CuSCN.[31-32] PEO of molecular weight 6×10$^5$ was purchased from Sigma Aldrich. For the preparation of the electrolytic solution, 0.5g of PEO was initially dissolved in 20 ml of acetonitrile under constant stirring for 3 hours. Then CuSCN was added to the above solution as weight fraction ($x$ wt.%) of PEO where $x = $ 0.25, 0.5, 1, 2, 3, and 5. The mixtures were stirred for another 21 hours to get a homogeneous solution.

**2.3 Fabrication of Cu-SPE based Memristive Cells**

The polymer electrolyte solution containing both PEO and CuSCN was spin-coated onto the ITO layer of the PET substrate. The spin-coating was carried out at 1000 rpm for 10 sec, followed by at 3000 rpm for 120 sec. The prepared films were then dried in a vacuum oven at 60 $^0$C for 5 hours to remove any left-out solvent. The thickness of the copper-ion conductive Cu-SPE film is estimated to be ~ 250 nm from the cross-sectional scanning electron microscope (SEM) image, as shown in supporting information (SI) Figure S5. Finally, a 40 nm thick circular copper (Cu) top electrode of diameter (~100 μm) was deposited, as mentioned in section 2.1. Hence, the desired device is a vertical stack of ITO/Cu-SPE/Cu.

**2.4 Characterization of Materials and Devices**

The structural properties and absorption spectra of the CuSCN film were characterized using X-ray diffraction (XRD) and UV-visible spectroscopy. The film's chemical structure and surface topography were studied using Raman spectra and atomic force microscopy (AFM). The Cu-SPE films were characterized by using XRD and FTIR. XRD study shows two broad peaks at 19.1° and 23.3° due to the PEO, as presented in SI Figure S3a. No Bragg peaks corresponding to CuSCN were observed in the electrolyte films. This confirms CuSCN is dissolved and ionic species are separated in the polymer electrolyte. SI Figure S3b shows the FTIR spectra of the Cu-SPE films with varying CuSCN concentrations. The characteristic vibrational modes of PEO and CuSCN are identified in the figure.

The electrical characterization of the CuSCN and Cu-SPE film-based memristive device were studied by using Keithley 4200-Semiconductor Characterization System (SCS). The PET substrate containing the vertical cells was placed on a probe station, and contact was made with the electrodes using the tungsten tips.

**3. RESULT AND DISCUSSION**

**3.1 Material Characterization**

The XRD spectra of as-received CuSCN powder and thin films spin-coated at 4000 rpm are shown in Figure 1a. As indexed, the powder CuSCN is observed to be in the β phase, which has a hexagonal (rhombohedral) structure.[33-34] After making films from the solution, the CuSCN though still remains in the β phase, becomes semicrystalline with a considerable amorphous fraction. The Bragg peaks observed at 16.27º and 27.4º are significantly broadened, as shown in the inset of Figure 1a. For the thin film, the average crystallite size for (003) and (101) peaks observed at 16.27º and 27.4º found to be ~ 9.33 nm and ~ 4.47 nm, respectively, as calculated using the Debye-Scherer formula. The UV-visible absorption spectra of the CuSCN powder and thin films deposited on ITO are presented in Figure 1b in the wavelength range of 200-800 nm. Above 350 nm (mainly in the visible region), no significant absorption is observed for the powder CuSCN. While the thin film of CuSCN shows some absorption in the visible region may be due to the defects or trap states created in the film.[35] The absorption spectra of the CuSCN powder and the thin film show a peak at ~ 297 nm, a characteristic of CuSCN.[35] The optical band gap energy was calculated from the Tauc plot, as shown in the inset of Figure 1b. The band gap energy of the CuSCN powder and thin film spin-coated at 4000 rpm were found to be 3.6 eV and 3.4 eV, respectively, which agrees well with the reported values.[36]

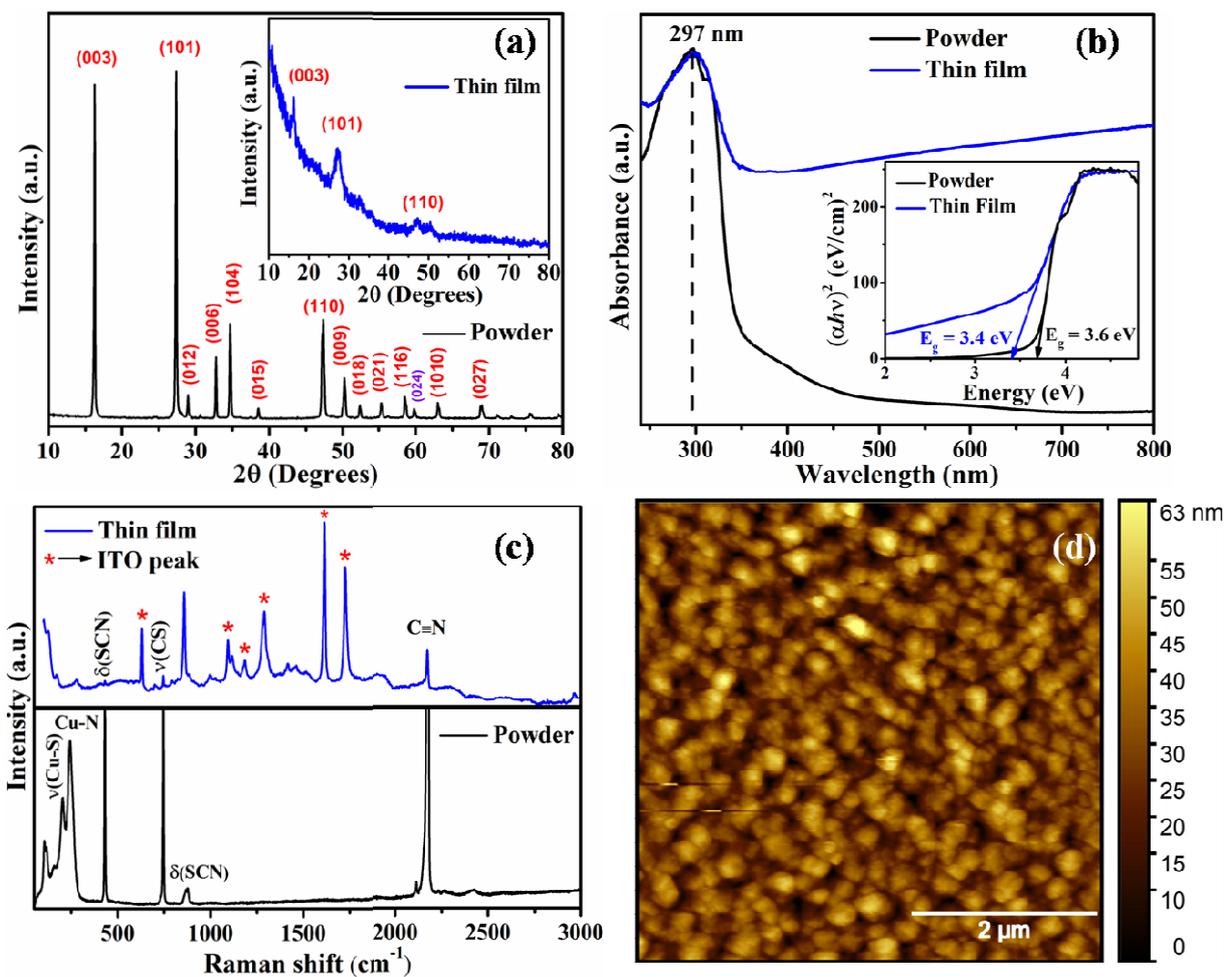

**Figure 1.** (a) XRD pattern of CuSCN powder and thin film (inset) deposited over ITO coated PET substrate, (b) UV-visible absorption spectra of CuSCN powder and thin film deposited over a glass substrate. Inset of (b) represents the determination of band gap energy of CuSCN powder and thin film from Tauc plot, (c) Raman spectra of CuSCN powder and thin film deposited over ITO coated PET substrate. The symbol star (*) represents the Raman peak of ITO, (d) AFM topography image of CuSCN thin film deposited over ITO coated PET substrate.

The Raman spectra in the frequency region of 50-3000 cm$^{-1}$ of the CuSCN powder and thin film are shown in Figure 1c. In the powder sample, the low-frequency modes at around 205 cm$^{-1}$ and 244 cm$^{-1}$ belong to the stretching vibration of Cu-S and Cu-N, respectively. The Raman shift at 430 cm$^{-1}$ and 876 cm$^{-1}$ corresponds to the bending vibration of S-C≡N, while the Raman shift at 746 cm$^{-1}$ is for C-S stretching. In the high-frequency region, CuSCN shows a single peak at ~ 2173 cm$^{-1}$ corresponding to the C≡N stretching.[24, 37] The thin film of CuSCN layer exhibits the peaks of S-C≡N bending vibration, C-S stretching mode, and C≡N stretching mode located at 430 cm$^{-1}$, 746 cm$^{-1}$, and 2173 cm$^{-1}$, respectively. These Raman peaks are characteristics of the β-phase of CuSCN. The Raman peaks indicated by the star mark (*) in the thin film correspond to the ITO layer on the PET substrate.[38] The surface topography image of CuSCN film deposited at 4000 rpm on ITO-coated PET substrate is presented in Figure 1d. The root mean square surface roughness measured over an area of 5μm × 5μm is ~ 10 nm with CuSCN grains on the surface.

3.2 **Memory Characteristics of CuSCN based memristive device**

The schematic diagram of the ITO/CuSCN/Cu devices as prepared by spin-coating of the CuSCN solution is presented in the inset of Figure 2a. Particularly for the device prepared at 4000 rpm while spin-coating, the current-voltage characteristics is shown in Figure 2a. The devices were biased over a small voltage range (0.8 V to – 0.8 V). As the voltage sweeps from 0V to 0.8 V, the current gradually rises, and the device attends some low resistance state (LRS). Again, by reverse basing (0 → – 0.8V → 0), the device steadily returns to the high resistance state. This *I-V* characteristic represents the ideal memristive behavior, *i.e.,* pinched hysteresis loop with zero crossing.[39] Such resistive switching memory characteristic is also known as the analog RS memory. The device shows very stable analog RS behavior over repeated cycles. Figure 2b shows the semilog plot of current vs. voltage for the first ten cycles of the voltage scan, where the analog RS behavior is almost reproducible without any significant deviation. The OFF-ON resistance ratio measured at a read voltage of 0.2V is 3.2 and 3.1 for the 1$^{st}$ and 10$^{th}$ cycle, respectively.

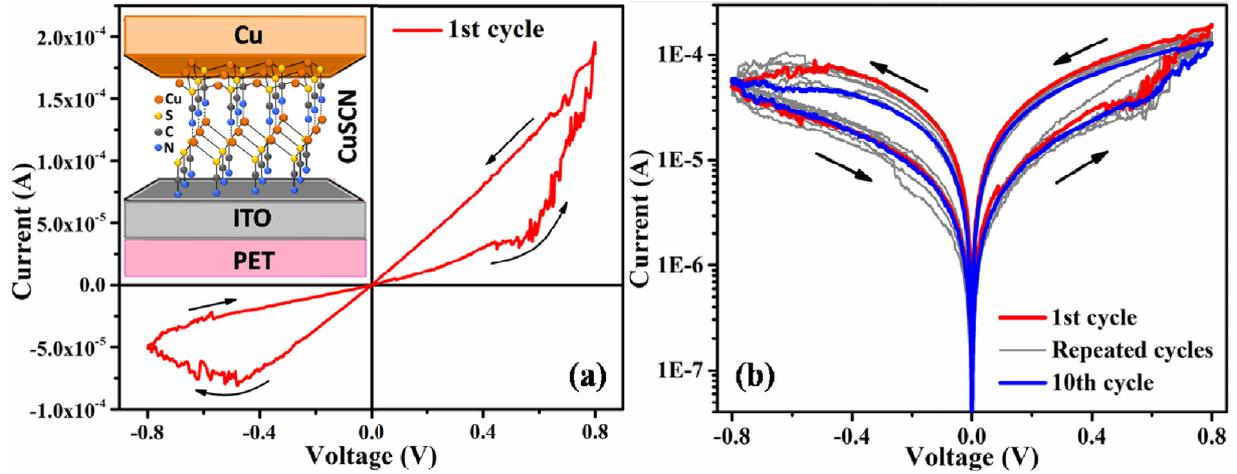

**Figure 2.** (a) Linear current-voltage (*I-V*) characteristic of ITO/CuSCN/Cu memristive device of first cycle by sweeping the voltage from 0 to +0.8V to – 0.8V to 0. The bias sweeping direction is indicated by black arrows. Inset of (a) shows the schematic device structure, (b) represents the semi-logarithmic plot of the memristive device in complete cycles by sweeping the voltage from 0 to +0.8V to – 0.8V to 0 for 10 continuous voltage sweep cycles.

As elucidated in the introduction, the gradual change in resistance or conductance state observed in the analog RS has an advantage over the digital RS memory for implementing neuromorphic computation. To further reveal the analog RS properties in this direction, we plotted *I-V* characteristics by sweeping the voltage for three different ranges ±0.5V, ±0.8V, and ±1.0V. In all these three sweeping cycles, the current gradually SETs to some LRS in the forward biasing direction ( 0 → 0.5V, 0.8V or 1.0V → 0 ) and RESETs to HRS progressively when biasing is reversed ( 0 → – 0.5V, – 0.8V or – 1.0V → 0 ). The LRSs obtained at a read voltage of 0.2V have resistance 110 kΩ, 43 kΩ, and 13 kΩ for voltage sweeping in the range of ±0.5V, ±0.8V, and ±1.0V, respectively. It is noteworthy that with increasing the voltage range of the sweeping cycle, the lower LRSs (or higher conductance states) are achieved while biasing in the forward direction. Hence as a consequence, in the reverse biasing direction also, HRSs are attained with less resistance as the voltage sweep range widens from ±0.5V to ±1.0V.

Another testimony of analog RS memory is depicted in Figure 3b. We successively bias the device in forwarding biasing direction with a sequence of 0 → 0.8V → 0 up to five consecutive cycles. After each cycle, the device achieved a new LRS state with lower resistance. Then in the reverse direction also, the device was biased with a sweeping sequence of 0 → – 0.8V → 0 consecutively for five cycles. The RESET process is also gradual, with an incremental rise in the resistance value of the HRS as the sweeping cycle varies from 6 to 10. This type of continuous and gradual change in the LRS and HRS states with bias modulating is the signature feature of analog RS memory.

To better represent the change in resistance states with voltage sweeps, Figure 3a and Figure 3b are reproduced in Figure 3c and Figure 3d, respectively, as the time evolution of current, as voltage sweeps. Figure 3c shows the steady rise in current as the range of sweeping voltage increases. Hence, multiple LRS and HRS can be achieved with different biasing voltages. In Figure 3d, repetitive forward biasing cycles produce incrementally higher conductance states, whereas the reverse biasing cycles produce progressively lower conductance states. These *I-V* characteristics relate to the synaptic functions, such as potentiation and depression of the synaptic weight.[9, 40-41] Hence, this is suggestive that the analog RS memory prepared with CuSCN has the propensity to show synaptic behavior and can be employed to implement neuromorphic computing. It is also observed in Figure 4a that the memory window (resistance ratio) increases with increase in the sweeping range of biasing voltage. Similarly, the memory window decreases with increase in cycle number in consecutive forward and reverse biasing cycles as shown in Figure 4b. In the forward biasing cycles, the current is self-limiting and may attain saturation after some cycles. In the reverse biasing cycles, the HRS state's resistance approaches the resistance of pristine cells as the number of cycles increases.

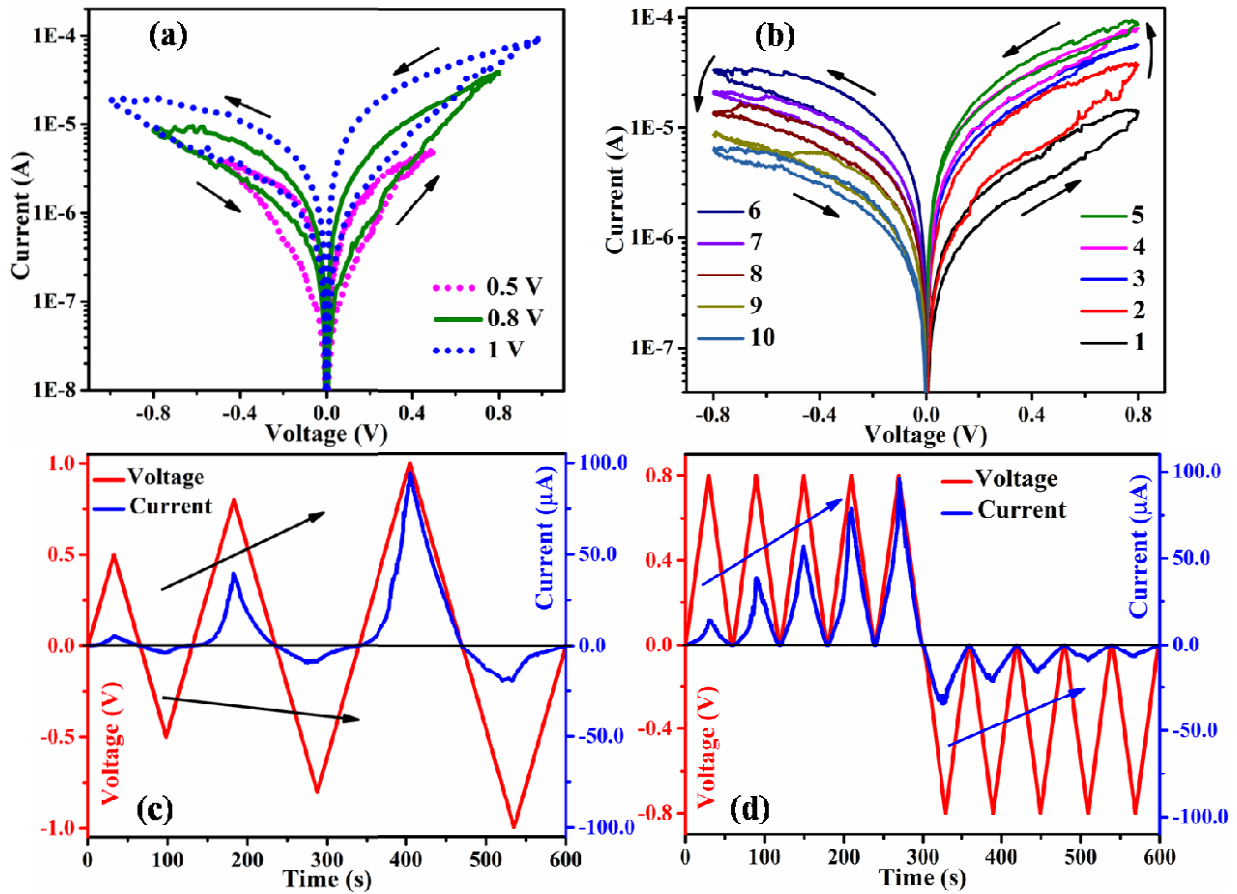

**Figure 3.** (a) Current-voltage characteristics of the ITO/CuSCN/Cu memristive device in cycles between 0V to ±0.5V (0V → 0.5V → 0V → − 0.5V → 0V), 0V to ±0.8V and 0V to ±1V. (b) Current-voltage characteristics of ITO/CuSCN/Cu memristive device, where 1, 2, 3, 4 and 5 represents the *I-V* curves under consecutive five positive voltage sweeps (0V → 0.8V → 0V) and 6, 7, 8, 9 and 10 represents *I-V* curves under consecutive five negative voltage sweeps (0V → − 0.8V → 0V), (c) Temporal evolution of current when voltage sweeps with widening range and (d) Temporal evolution of current for five consecutive positive and negative voltage sweeps.

The memristive device ITO/CuSCN/Cu can also show digital RS memory after a forming stage which occurs just above 1.0 V. But in successive cycles, SET and RESET occur at very low voltages ~ 0.4V and ~ − 0.25V, respectively, as shown in Figure 5a. Figure 5b shows the OFF-ON resistance ratio observed for 40 cycles as ~$10^5$. Very few reports on memristive systems showed digital and analog RS memories together.[9-11] The simultaneous observation of digital and analog RS memory makes the CuSCN-based devices more versatile. To gain more insight into the RS behavior of CuSCN-based devices, we carried out the electrical characterization of ITO/CuSCN/Cu devices with different CuSCN layer thicknesses prepared at 3000 and 5000 rpm of spin-coating. The devices prepared at 5000 rpm were found to be mostly short-circuited and hence discarded from further investigation. The devices prepared at 3000 rpm don't show any analog RS behavior, as shown in supporting information Figure S1a. However, it shows stable bipolar digital RS behavior when swept between 0 and ±2V (SI Figure S1b). Compared to the devices prepared at 4000 rpm, the SET voltage is higher. The observation of both analog and digital RS in devices prepared at 4000 rpm and the absence of analog RS switching in devices prepared at 3000 rpm suggests that the thickness of the CuSCN layer is the crucial factor in observing the ideal memristive behavior. To ascertain whether the resistive switching phenomena is due to the electrochemical metallization (ECM) (by copper ion diffusion from the top Cu electrode) or trapping of charge carriers at the interface, we prepared the devices with a gold top electrode (ITO/CuSCN/Au). The thickness of the CuSCN layer is maintained the same by spin-coating it at 4000 rpm. The device shows both the analog and digital RS memory (SI Figure S2a,b), similar to the ITO/CuSCN/Cu device. This suggests that the RS phenomenon is due to charge trapping and de-trapping at the interface. For such interfacial RS behavior, high electronic conductivity (~ 0.01 S m$^{-1}$) of the CuSCN layer owing to the hole transport plays a pivotal role.[42] For materials like MXene ($Ti_3C_2$) and 1T-$MoS_2$ nanosheets, the role of electronic conductivity (due to electron or hole) in observing RS is already established.[10, 39]

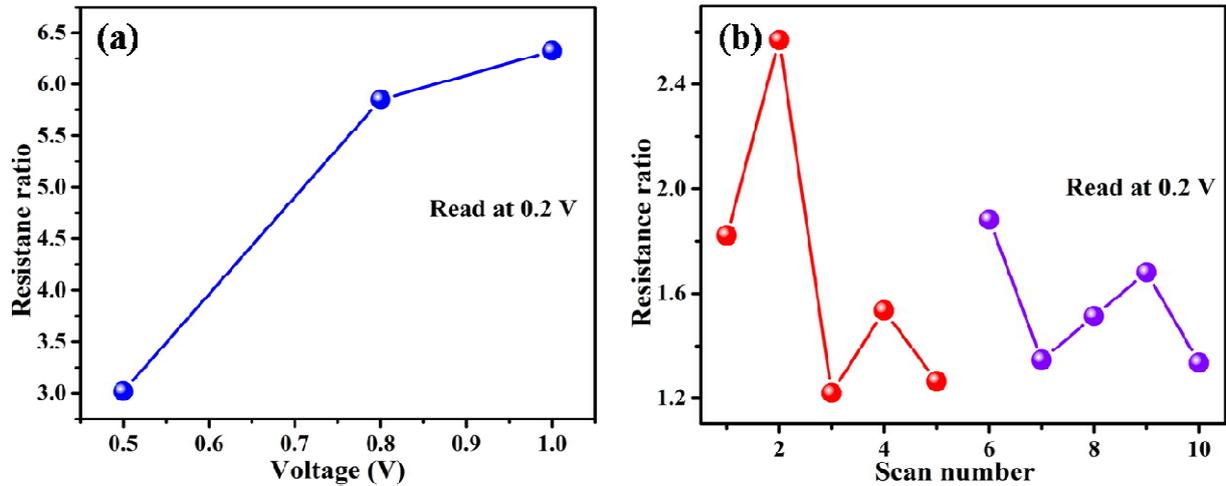

**Figure 4.** (a) Plot of resistance ratio for different voltage sweeping cycles, (b) Plot of resistance ratio with consecutive positive and negative voltage sweeping.

Although we are ruling out the possibility of an ECM type of mechanism for RS in CuSCN, the role of Copper ion ($Cu^+$) diffusion can't be completely played down if we are looking at the voltage range over which the RS is observed in these devices (with Au and Cu as a top electrode). Unlike the case of the copper top electrode, we could not observe the analog switching for lower voltage sweep (< 1.0V) with the gold top electrode. Similarly, the SET and RESET voltages of digital RS memory in gold-based devices are much higher than in copper-based devices. So, we believe the RS memory behavior in ITO/CuSCN/Cu devices is charge carrier trapping and de-trapping modulated by the copper ion diffusion from the top copper electrode. Such role of ionic modulation and coupling of ionic and electronic currents in governing the resistive switching behavior is already testified in other memristive systems.[43-45]

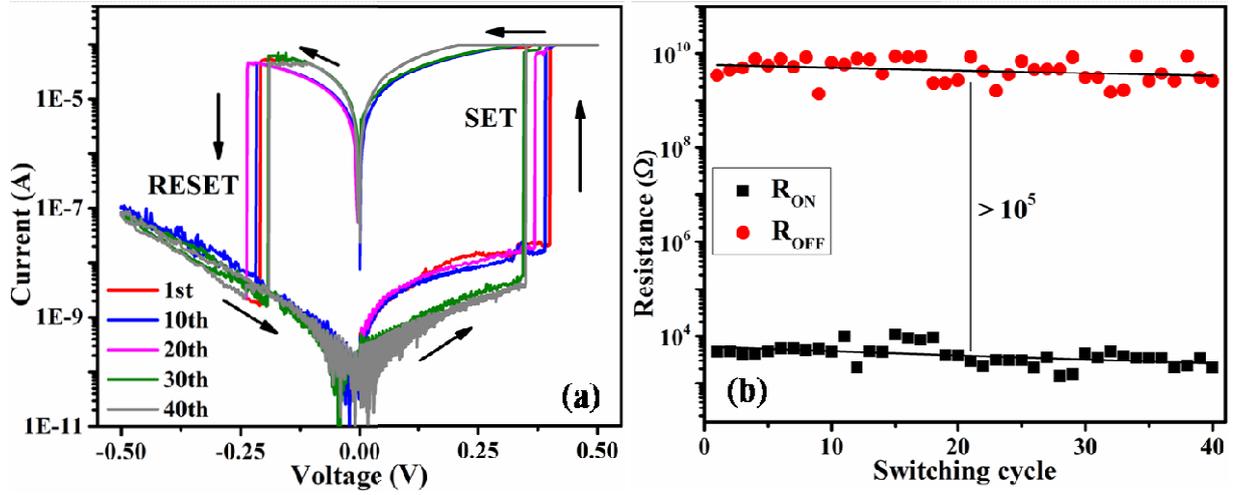

**Figure 5.** (a) Bipolar *I-V* characteristics of ITO/CuSCN/Cu memristive device, (b) Plot of ON and OFF state resistances with switching cycle.

**3.3 Current conduction mechanism in CuSCN based memristive devices**

To elucidate further the current conduction mechanism in the ITO/CuSCN/Cu devices, the analog *I-V* characteristics for the first positive voltage scan (Figure 3b) is plotted in log-log scale as shown in Figure 6a. It is observed that in the low biasing region, i.e., 0 < V < 0.27 V, the slope of the linearly fitted line is nearly one, indicating the Ohmic conduction due to the intrinsic charge carriers in the CuSCN layer. The non-linear part of the curve of Figure 6a in the voltage region 0.27 V < V < 0.5 V is plotted as ln(*I*) vs $\sqrt{V}$ as shown in the inset of Figure 6a. The curve is well fitted by the linear relationship, confirming that the current conduction is governed by Schottky emission in the particular voltage region. The slope of the curve in the biasing region 0.5V < V < 0.8 V is 2.96. It suggests that the current conduction in the higher voltage region is the trap-assisted space-charge-limited conduction (SCLC), *i.e.,* I α $V^m$ where m > 2.[9, 46-48] Thus, in the HRS, the current conduction is Ohmic initially and then followed by Schottky emission and trap assisted-SCLC. After the SET process, when sweeping back to zero from 0.8 V, the slope is 1.02 for the entire biasing region, indicating the Ohmic type conduction again in the LRS. In the negative sweeping direction, the current conduction follows the same sequence of Ohmic, Schottky emission, and SCLC while sweeping the bias from 0 V to – 0.8V.

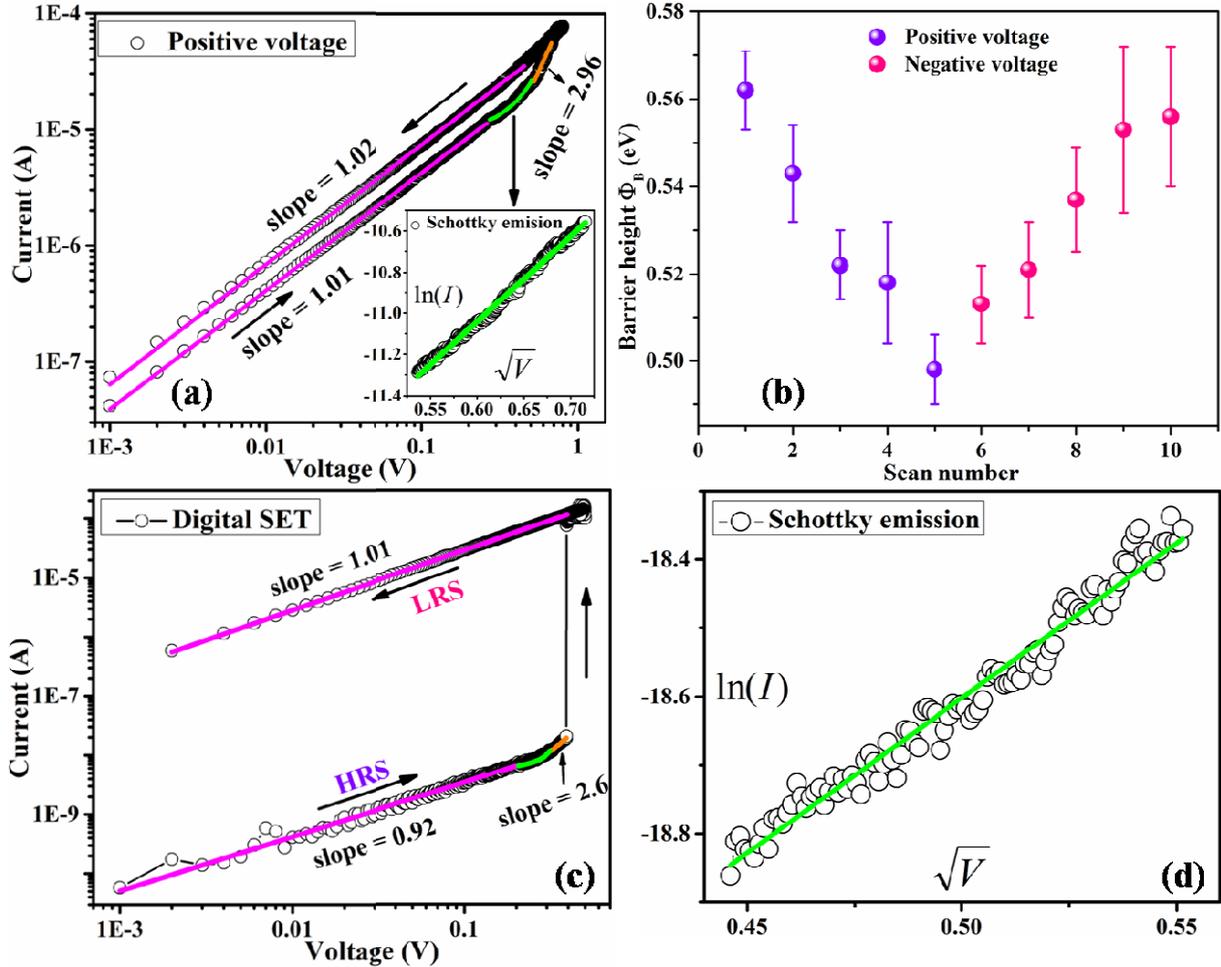

**Figure 6.** (a) log-log plot of the analog *I-V* for the first positive voltage scan of the ITO/CuSCN/Cu memristive device. The inset of (a) represents the ln(*I*) vs $\sqrt{V}$ plot with linear fitting in the voltage region 0.27 V< V < 0.5 V, (b) Plot of Schottky barrier height ($\phi_B$) with consecutive positive and negative voltage scan, (c) log-log plot of the digital *I-V* for the first SET process in the ITO/CuSCN/Cu memristive device, (d) ln (*I*) vs $\sqrt{V}$ plot of Fig. (c) with linear fitting in the voltage region 0.20 V < V < 0.31 V.

CuSCN is a p-type semiconductor with work function ($\Phi_{CuSCN}$ ~ 5 eV).[49] Hence, with Cu ($\Phi_{Cu}$ = 4.4 eV) and ITO ($\Phi_{ITO}$ ~ 4.8 to 5.2 eV), CuSCN makes Schottky and Ohmic contacts, respectively, at both the interfaces.[50-52] So, during positive and negative voltage sweeping cycles, the Schottky contacts get modified, influencing the charge flow to and from the device. At the Cu/CuSCN interface, the Schottky barrier height can be calculated by using the Richardson-Schottky equation. The Richardson-Schottky equation of the *I-V* characteristics is given by

$$I = AA^*T^2 e^{\{-q(\varphi_B - \sqrt{qE/4\pi\varepsilon_i})/kT\}} \tag{1}$$

where, $A = \pi(50\ \mu m)^2$ = cell contact area, $A^* = \frac{4\pi q m_n^* k^2}{h^3}$ = Richardson constant, $T$ = temperature, $q$ = electric charge, $\varphi_B$ = Schottky barrier height, E = electric field, $\varepsilon_i$ = dielectric constant of material and k = Boltzmann constant.[9] Taking the natural logarithm of equation (1) and replacing *E* by *V/d*, where *d* is the distance between the Cu top and ITO bottom electrode, we get

$$ln(I) = \left\{ln(AA^*T^2) - \frac{q}{kT}\varphi_B\right\} + \frac{q}{kT}\sqrt{q/4\pi\varepsilon_i}\ \sqrt{V} \tag{2}$$

By taking the value of A* = 119.56 A/K²cm² ($m_n^* = m_0$), the Schottky barrier height ($\varphi_B$) is calculated from the intercept of $ln(I)$ vs $\sqrt{V}$. We find that the Schottky barrier decreases under successive positive voltage scans and increases under subsequent repetitive negative voltage scans, as shown in Figure 6b. Further, from the slope of $ln(I)$ vs $\sqrt{V}$ plot, the ideality factor is calculated to be ≈ 10, much greater than the ideality factor of an ideal diode (n =1). The higher ideality factor explains the inhomogeneity in the Schottky barrier that arises due to traps or interfacial states at the Cu/CuSCN interface.[53] Quite a similar conduction mechanism is observed for digital RS memory, as shown in Figure 6c. The log-log plot of current-voltage characteristics during the SET process shows linear fit in the low voltage region of HRS with slope ~ 0.92. In the high voltage region, i.e., 0.21 V to 0.39 V, the current conduction mechanism follows Schottky emission, as shown in Figure 6d, followed by trap-assisted SCLC. In the LRS, the slope is well-fitted by Ohmic conduction.

Based on the conduction mechanism discussed, the energy band diagram is schematically presented in Figure 7a-d. In the ITO/CuSCN/Cu memristive device, the Cu/CuSCN and ITO/CuSCN interfaces make Schottky and Ohmic contacts (Figure 7b). When a positive bias is applied to the Cu top electrode, band bending occurs, and as a result, Schottky barrier height reduces. Hence, the current flow gets easier, and the trap states get filled. This makes the device ON. On the other hand, in the reverse polarity of the bias, the Schottky barrier height increases at the Cu/CuSCN interface, and the current flow reduces. The filled trap states release the charge carriers at the Ohmic contact made at the ITO/CuSCN interface. Hence, charge trapping and de-trapping at the interfacial states is the main reason behind analog RS behavior in ITO/CuSCN/Cu memristive devices. The electrochemical oxidation of the Cu electrode and influx of $Cu^+$ ions into the CuSCN medium during the forward biasing conditions can also strongly influence the switching mechanism. Particularly in the digital RS memory, the ON state is due to the conductive filament formed as a result of both filled trap states and the reduction of Cu+ ions to metallic copper-dendrites grown at the ITO/CuSCN interface. The conductive filament may not be entirely due to charge trapping or the copper dendrites connecting both interfaces. Instead, it can be the result of both phenomena. A percolative conductive path may be formed between filled interfacial trap states with patches of copper-dendrites in the CuSCN layer. Hence, the off state may be attained by partial dissolution of the conductive filament due to the offloading of charge carriers from the trap-states and oxidation of copper-dendrites.

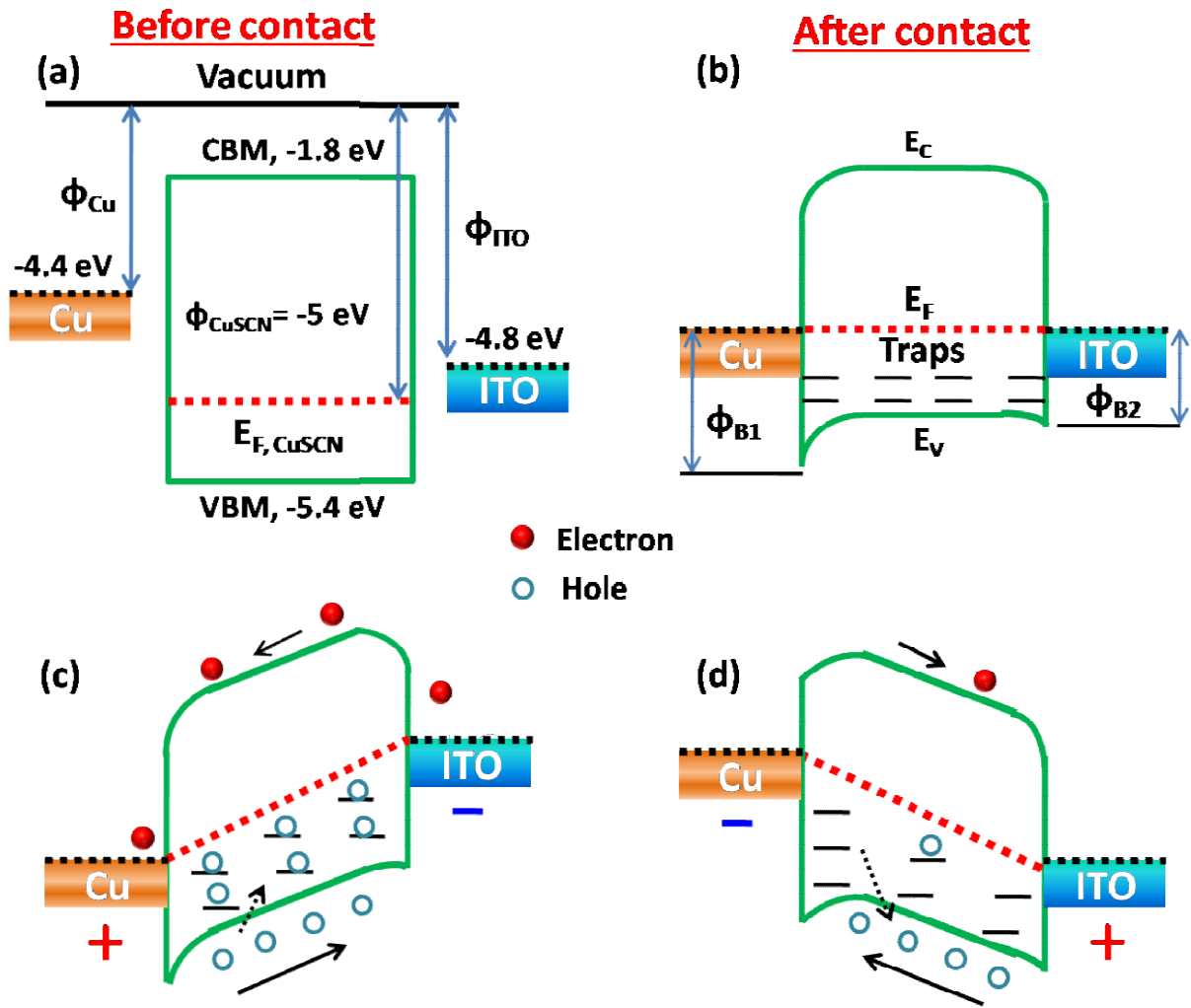

**Figure 7.** (a) Energy band diagram of the ITO/CuSCN/Cu memristive device before contact, (b) after contact, (c) during the transition from OFF state to ON state (SET process), and (d) ON state to OFF state (RESET process).

**3.4 Current-voltage characteristics in ITO/Cu-SPE/Cu memristive device**

Cu-SPE is the electrolyte medium containing CuSCN in the ionized form as $Cu^+$ and $SCN^-$ ions embedded in the PEO matrix, as shown in the schematic diagram in Figure 8a. The structural and spectroscopic characterization of the Cu-SPE film is already presented in the supporting information (Figure S3), confirming the dissolution of CuSCN in the PEO. To optimize the CuSCN concentration in Cu-SPE for better RS characteristics, the ITO/Cu-SPE/Cu cells were prepared with Cu-SPE containing $x$ wt.% of CuSCN where $x$ = 0.25, 0.5, 1, 2, 3, and 5. The *I-V* characteristic of ITO/Cu-SPE/Cu cells containing varying CuSCN concentrations (0.25% to 5.0%) are shown in the SI Figure S6. All the devices show bipolar digital RS behavior except the one with 5 wt.% of CuSCN. However, the RS behavior is unstable and not repeatedly observed for 0.25 and 0.5 wt.%. Whereas above 1 wt.%, the SET, and RESET processes occur at higher positive and negative biases with increasing concentrations of CuSCN. The concentration of the ionic moieties ($Cu^+$ and $SCN^-$) increase along with CuSCN concentration in the Cu-SPE film. Hence, the depolarizing field created due to the space-charge polarization becomes more dominating in Cu-SPE film containing higher concentrations of CuSCN. Due to this, SET and RESET voltages increase in ITO/Cu-SPE/Cu devices with higher CuSCN concentration, and at 5wt.%, the device could not be SET even if the applied bias is increased up to 10.0V. This is commonly observed in polymer electrolyte-based ECM-type memristive devices.[54] For further characterization of ITO/Cu-SPE/Cu devices, we fixed the CuSCN concentration in Cu-SPE as 1.0 wt.%, which shows very stable and repetitive bipolar digital RS behavior.

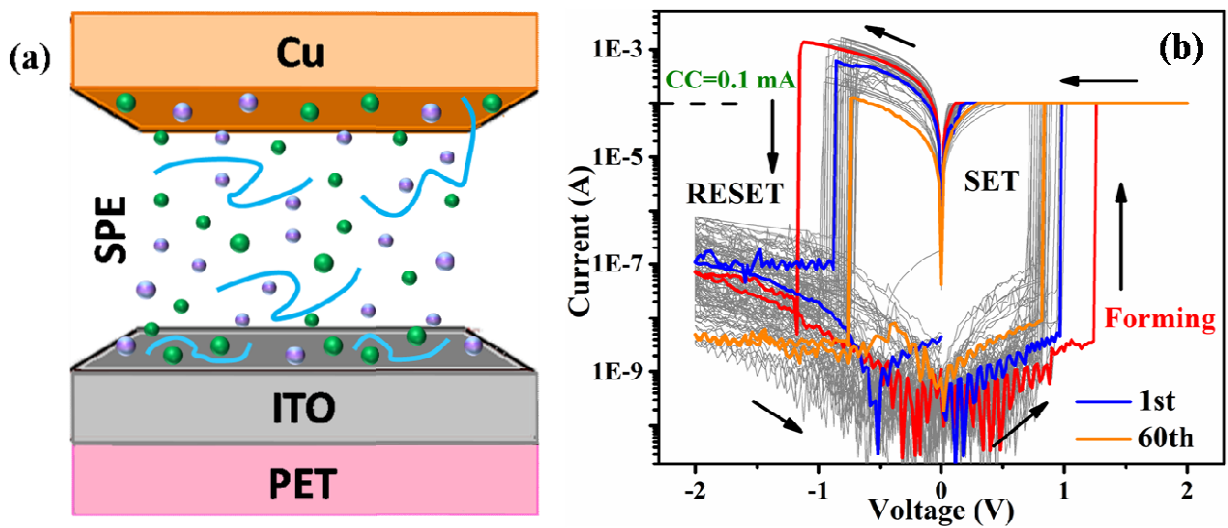

**Figure 8.** (a) Schematic diagram of the ITO/Cu-SPE/Cu memristive device, (b) Bipolar *I-V* characteristics of the ITO/Cu-SPE/Cu memristive device for 1 wt% CuSCN.

Figure 8b represents the *I-V* characteristics of ITO/Cu-SPE/Cu devices containing the Cu-SPE film with 1 wt% of CuSCN for 60 consecutive voltage sweep cycles. In the first sweep cycle, the device was SET at a higher voltage of 1.24V, called the forming process, and RESET at –1.16V. In the second sweep cycle, the device was SET at 0.96V and RESET at – 0.86V. The *I-V* plot shows digital RS memory with no significant variation of SET and RESET voltages from cycle to cycle.

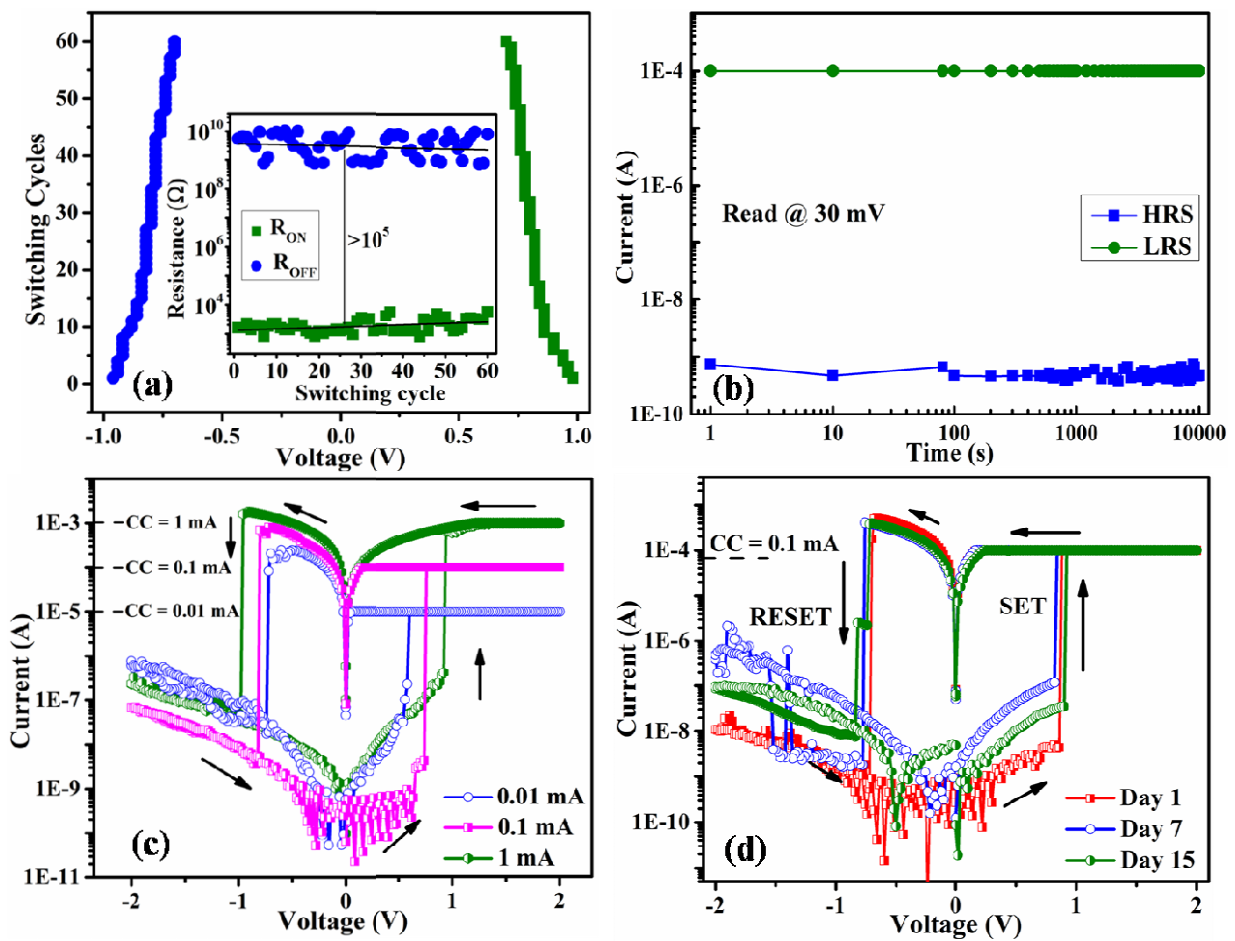

**Figure 9.** (a) Distribution of the SET and RESET voltage of ITO/Cu-SPE/Cu memristive device. The inset of (a) shows the ON and OFF state resistances with switching cycles (b) Retention test of the ON and OFF states at 30 mV read voltage (c) Multi-level resistive switching observed in the ITO/Cu-SPE/Cu memristive device with CuSCN concentration of 1 wt% (d) Current-voltage characteristics of the device up to 15 days.

Further, Figure 9a represents the SET and RESET voltages distribution of ITO/Cu-SPE/Cu cells during 60 consecutive voltage sweep cycles. The average SET and RESET voltages are 0.80 V and –0.78 V. The ITO/Cu-SPE/Cu cells exhibit good cyclic behavior, as shown in the inset of Figure 9a. The ON and OFF states are reproducibly achieved for the SET and RESET cycles. The OFF-ON state resistance ratio is more than $10^5$, maintained throughout all the measuring cycles. Also, retention characteristic was measured in each particular state under continuous 30 mV read voltage, as shown in Figure 9b. Both ON and OFF states are stable up to the measured time duration, i.e., $10^4$ sec. The Cu-SPE based cells also show multi-level bipolar resistance switching, as presented in Figure 9c. Three resistance states were achieved by keeping the compliance at 0.01 mA, 0.1 mA, and 1.0 mA. As the compliance current increases from 0.01 mA to 1 mA, the SET and RESET voltage increase. The ON state resistances are in the order of 10 kΩ, 2 kΩ, and 500 Ω for 0.01 mA, 0.1 mA, and 1.0 mA compliance, respectively. This shows that with increasing compliance current, the ON-state resistance decreases steadily. The *I-V* measurements of the device for up to 15 days with an interval of 7 days are shown in Figure 9d. The device showed reliable and reproducible resistive switching behavior over the period without any deterioration of the switching characteristics as an effect of storing or aging.

### 3.5 Current conduction mechanism in Cu-SPE based memristive devices

Further, to understand the role of the top Cu electrode and its electrochemical effect on the switching mechanism, a memristive device was prepared with Au inert electrode by replacing the copper. In the ITO/Cu-SPE/Au memristive device, no switching is observed up to the applied bias of ±5 V, as shown in SI Figure S7. The current remains very low (~ 1.23 μA) even at 5V. This suggests that the Cu top electrode has a definite role in the resistive switching behavior of ITO/Cu-SPE/Cu memristive devices. To investigate the switching mechanism in Cu-SPE, the *I-V* characteristic of one switching cycle was plotted in a log-log scale, as shown in Figure 10a for the SET process. The current conduction mechanism in the bias region 0.1V < V < 0.78V is observed to be linear with a slope of 1.46. This can be understood using Mott-Gurney equation for ion-hopping as given in equation (3). The Mott-Gurney equation is given by[55]

$$I = 2zecav\, exp\left(\frac{W_a^0}{kT}\right) sinh\left(\frac{azeE}{2kT}\right) \quad (3)$$

where, $c$ is the concentration of mobile cations, $a$ is the jump distance of ions, and $v$ is the frequency factor. For low electric fields $\left(E < \frac{kT}{aze}\right)$, the equation (3) can be modified as shown in equation (4) with linear dependence of $I$ on $E$ similar to Ohmic conduction.

$$I = \frac{(ze)^2 cE}{kT} a^2 v\, exp\left(\frac{W_a^0}{kT}\right) \quad (4)$$

This linear behavior infers that the $Cu^+$ ions drifted toward the ITO electrode in the electrolyte medium to form copper filaments on applying a positive bias to Cu top electrode.

In the LRS, the slope is found to be 1, which means that the current conduction mechanism was dominated by Ohmic conduction as copper filaments are formed in the Cu-SPE film between Cu and ITO electrodes. During the RESET process in the LRS, from 0 to RESET voltage (– 0.84V), the slope is 1 indicating the Ohmic type conduction. In the HRS from – 0.84V < V < – 0.16V, the current conduction mechanism follows space charge limited current (SCLC) with a slope of 2.54.[45] This could be due to the emergence of space charge in the electrolyte medium immediately after the breakdown of the filament. As the voltage decreases further, the current follows linearly with the voltage due to the ion-hopping in the electrolyte medium, as explained by equation (4). These studies confirm the switching mechanism to be ECM or conductive-bridge random access memory (CBRAM) type, as observed in other polymer-electrolyte systems with electrochemically active electrodes.[56,57] The formation and dissolution of copper filament are responsible for the ON and OFF states, and are controlled by the electrochemical redox reaction at the interfaces.

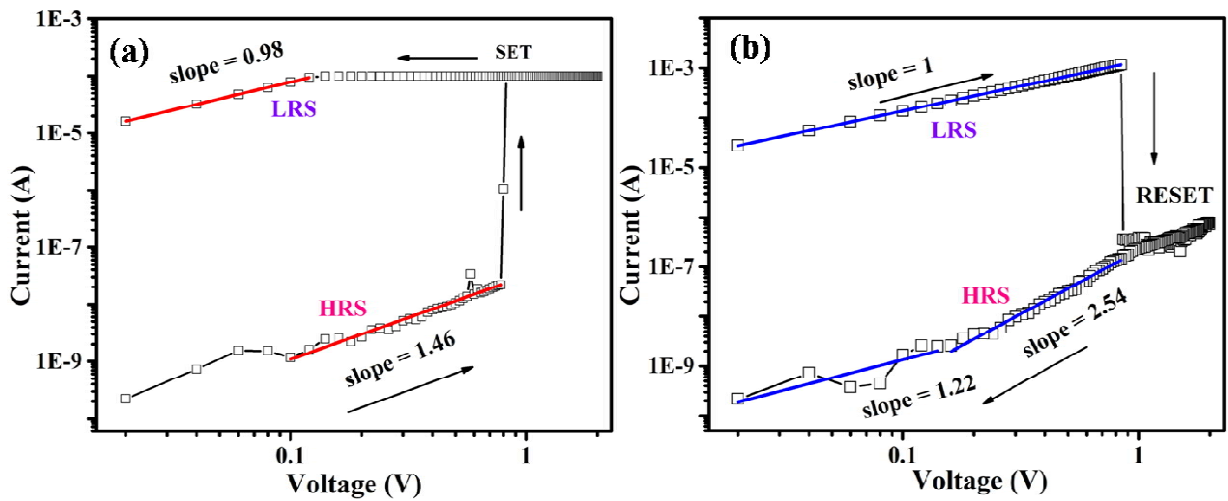

**Figure 10.** Linear fitting of the *I-V* curve for the (a) SET and (b) RESET process using log-log plot for CuSCN concentration of 1 wt% in ITO/Cu-SPE/Cu device.

Moreover, no analog RS is observed in these ECM-type memory devices based on Cu-SPE film. In fact, analog RS is less often observed in ECM-type memory devices and more rarely so in polymer-electrolyte based ECM devices.[41,58] If the switching layer is electronically insulator, sharp SET and RESET process occurs, leading to digital RS. But when the switching layer is semiconducting (due to electron or hole conduction) along with allowing ion transport, the chances of analog RS arises.[59]

## 4. CONCLUSIONS

In conclusion, we prepared and studied two types of memristive devices based on CuSCN. The switching media in these two devices are CuSCN and an SPE where CuSCN remains dissolved in the PEO matrix (Cu-SPE). While the CuSCN is known to be a p-type semiconductor, the Cu-SPE is an ionic conductor and electronically insulator. The first device consisted of a solution-processed CuSCN layer as the switching medium. The RS behavior is strongly dependent on the thickness of the CuSCN layer. Devices with thinner CuSCN layer (prepared at 4000 rpm during spin-coating) show both analog and digital RS memory. The switching mechanism is due to the trapping and de-trapping of charge carriers at the trap or interfacial states modulated by the Schottky emission. The copper-ion diffusion from the top Copper electrode also influences the RS characteristics. The second device, made up of the Cu-SPE layer, shows only digital RS memory. The devices offer good cyclability and retention of ON and OFF states. Multi-level RS switching is also possible, as demonstrated by achieving discrete low resistance states (LRSs) by enforcing different current compliances. However, analog RS is not observed in Cu-SPE-based devices. The switching mechanism in the Cu-SPE based devices is ECM type, where the formation and dissolution of copper filament are responsible for resistive switching behavior. Nevertheless, CuSCN and its electrolytic system show versatile and contrasting RS memory characteristics. Hence, they have immense potential for developing low-cost solution-processable nanoelectronics for digital non-volatile memory and neuromorphic computing applications.

**ACKNOWLEDGMENTS**

The authors gratefully acknowledge the fund received from the Department of Science and Technology, Government of India, through the DST-FIST project (SR/FST/PSI-212/2016(C)). The authors also sincerely thank the help received from MRC, MNIT Jaipur in RAMAN and AFM measurements.

# Supporting information

Contrasting Analog and Digital Resistive Switching Memory Characteristics in Solution-Processed Copper (I) Thiocyanate and Its Polymer Electrolyte Based Memristive Devices

*Rajesh Deb[1], Saumya R. Mohapatra\*[1], Manjula G. Nair[2], and Ujjal Das[3]*

[1]Solid State Ionics Laboratory, Department of Physics, National Institute of Technology Silchar, Silchar-788010, Assam, India

[2]Department of Physics, Indian Institute of Technology, Patna, Bihar, India, 801106

[3]Quantum Materials and Device Unit, Institute of Nano Science and Technology, Mohali-140306, Punjab, India

*Email:saumya@phy.nits.ac.in

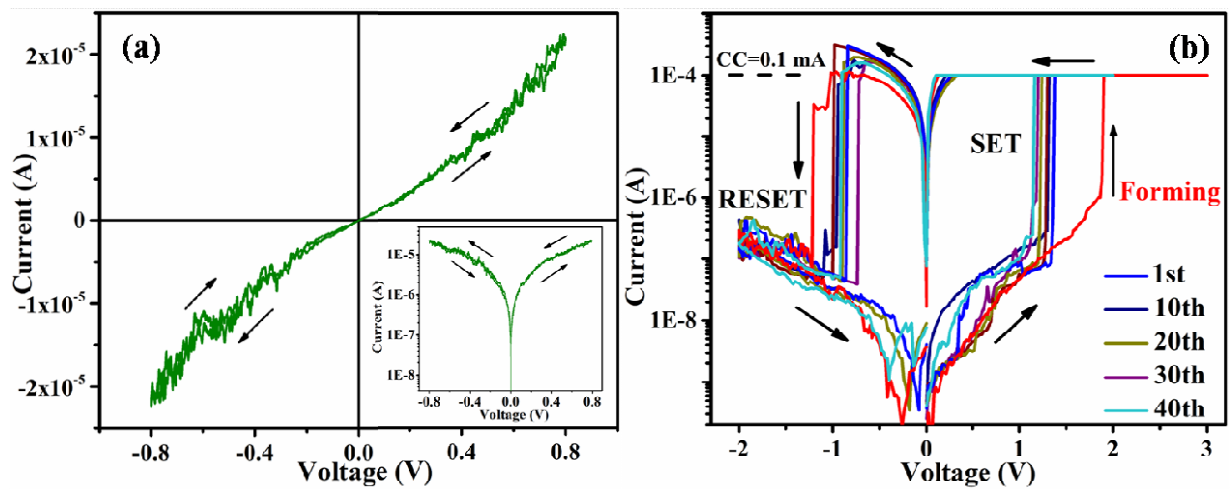

**Figure S1.** (a) Current-voltage characteristics of the Cu/CuSCN/ITO memristive device spin-coated at 3000 rpm. Inset of (a) represents the semi-logarithimic plot. (b) Bipolar *I-V* curves of Cu/CuSCN/ITO memristive cell spin coated at 3000 rpm.

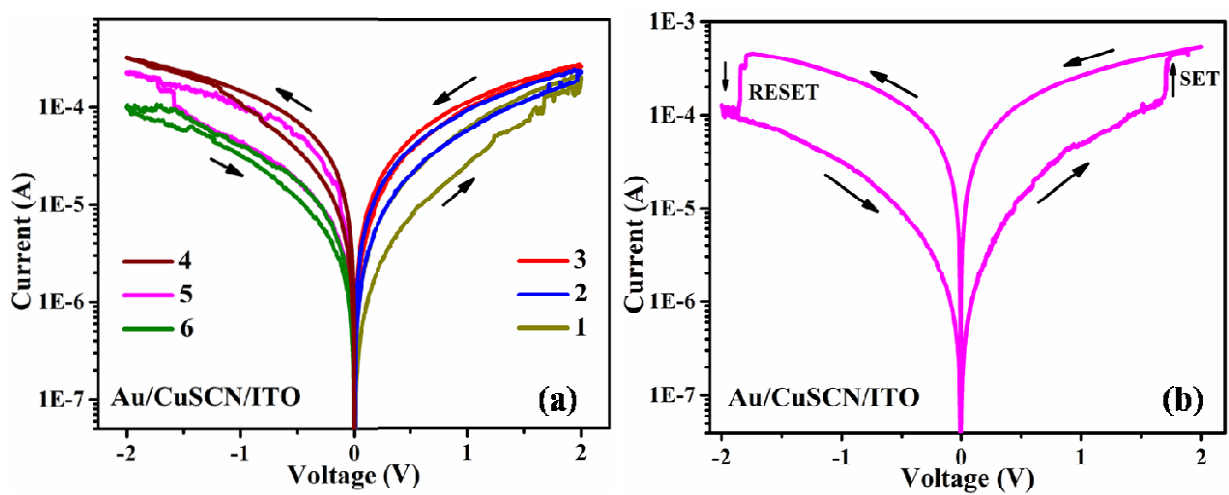

**Figure S2.** (a) Current-voltage characteristics of Au/CuSCN/ITO memristive device, where 1, 2, 3 represents the *I-V* curves under consecutive three positive voltage sweeps (0V → 2V → 0V) and 4, 5, 6 represents *I-V* curves under consecutive three negative voltage sweeps (0V → – 2V → 0V), respectively (b) Bipolar digital resistive switching as observed from the *I-V* characteristics in the Au/CuSCN/ITO memristive device.

## Structural and Optical characterization of Cu-SPE

The X-ray diffraction pattern of the copper ion conductive solid polymer electrolyte films with CUSCN (Cu-SPE) concentration of 0.25% to 5% is shown in Figure S3(a). The pattern shows two dominant peaks of PEO at 19.1° and 23.3°, which corresponds to (120) and (112) planes respectively.[1,2] It is observed that there is an increase in the intensity of (120) peaks with the rise in the salt concentration. This suggests that salt concentration has a strong influence on the crystallization process of the PEO. Further, no Bragg peak corresponding to CuSCN is observed in the electrolyte film, even at 5 wt.% of the CuSCN concentration. This indicates that no uncomplexed CuSCN is present in the electrolyte film. The FTIR spectra of the Cu-SPE film plotted as transmittance vs. wavenumber (580-3500 cm$^{-1}$) is shown in Figure S3(b). It confirms the characteristic vibrational modes of PEO in the wavenumber region 800-3000 cm$^{-1}$. The bands at 841cm$^{-1}$ and 946 cm$^{-1}$ are due to CH$_2$ asymmetric rocking motion with some contribution of C-O stretching. The sharp band at 1094 cm$^{-1}$ is assigned to the C-O-C stretching mode. The relatively small bands at 1279 cm$^{-1}$, 1343 cm$^{-1}$, and 1463 cm$^{-1}$ are because of CH$_2$ twisting, CH$_2$ wagging, and CH$_2$ deformation, respectively. The absorption band located at 2880 cm$^{-1}$ is assigned to the C-H asymmetric stretching mode.[3,4] The C-N stretching band found at 2168 cm$^{-1}$ was visible for higher concentrations of copper (I) thiocyanate salt.

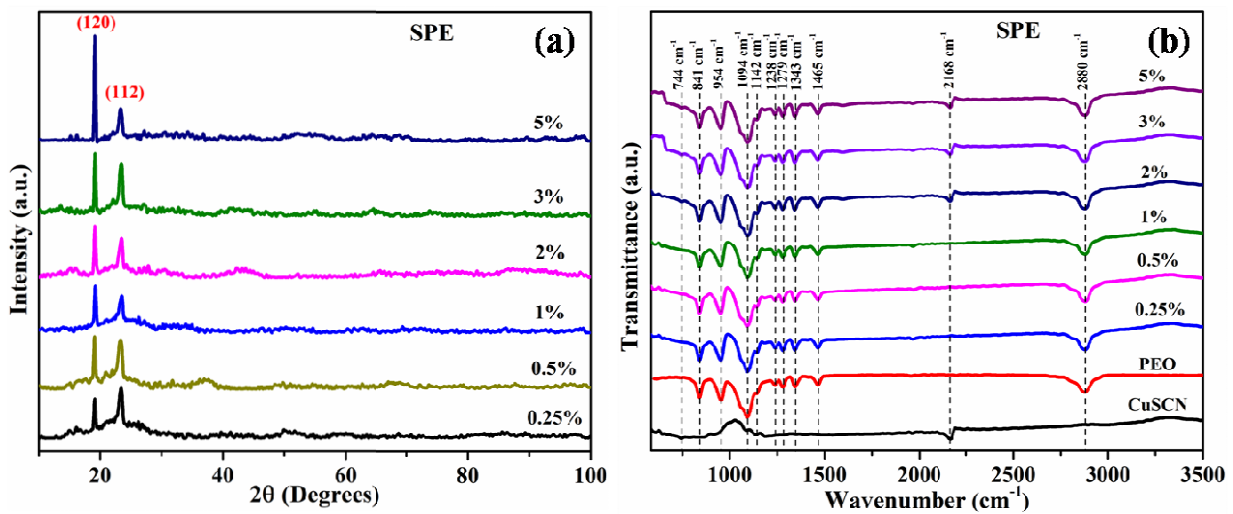

**Figure S3.** (a) X-ray diffraction pattern of Cu-SPE films, (b) FTIR spectra of Cu-SPE film in the wavenumber region 580-3500 cm$^{-1}$.

# X-ray Photoelectron Spectroscopy (XPS) of Cu-SPE

The chemical and electronic state of atoms in the electrolyte film was investigated by X-ray photoelectron spectroscopy (XPS). The XPS full scan spectrum as shown in Fig. 10 (a), indicates the presence of Cu, S, C, N, and O in the film. The high-resolution spectrums of the S 2p, C 1s, N 1s, O 1s, and Cu 2p are shown in Fig. S4 (b)-(f). In the S 2p core level spectra, the component peak observed at binding energy (BE) of 162.2 eV is due to either C-S or Cu-S. The other two peaks at 163.8 eV and 168.8 eV correspond to S-C≡N and S-O environments. The C 1s core level spectra is dominated by a peak at 285.9 eV, which corresponds to CuSCN (S-C≡N), and the peak at 284.4 eV is due to the C-H environments of PEO. N 1s spectra show a dominant peak of

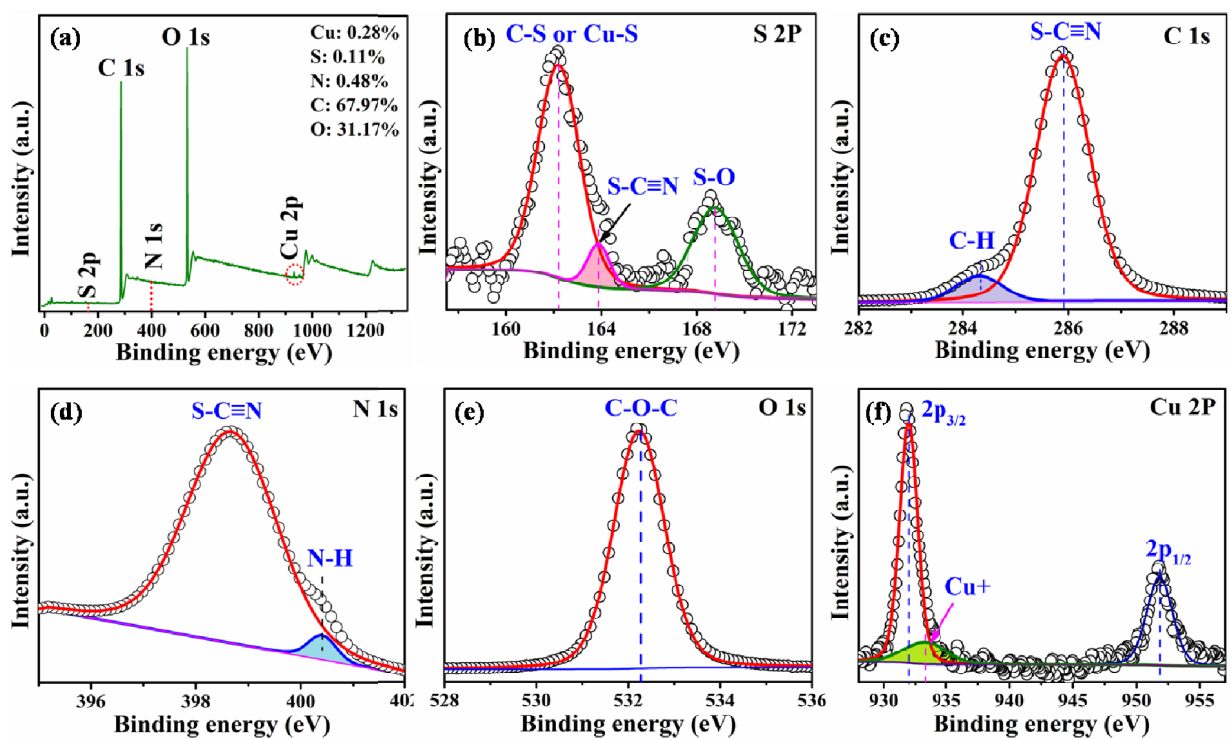

**Figure S4.** (a) XPS survey spectrum of the Cu-SPE film for 1 wt% of CuSCN. (b)-(f) High resolution spectra of the S 2p, C 1s, N 1s, O 1s and Cu 2p regions of the Cu-SPE film.

CuSCN at 398.6 eV, and a small peak at 400.3 eV is due to the N-H environment. The O 1s spectrum of the electrolyte film consists of one peak at 532.3 eV, assigned as C-O-C. From the high-resolution spectra of the Cu 2p core levels, the peaks at 932 eV and 951.8 eV correspond to the prominent peaks of Cu $2p_{3/2}$ and Cu $2p_{1/2}$, respectively. The component peak at 933.1 eV is due to the presence of $Cu^+$ in the electrolyte film.[5-10] These sources indicate that the copper (I) thiocyanate separated into $Cu^+$ and $SCN^-$ ions in the polymer electrolyte.

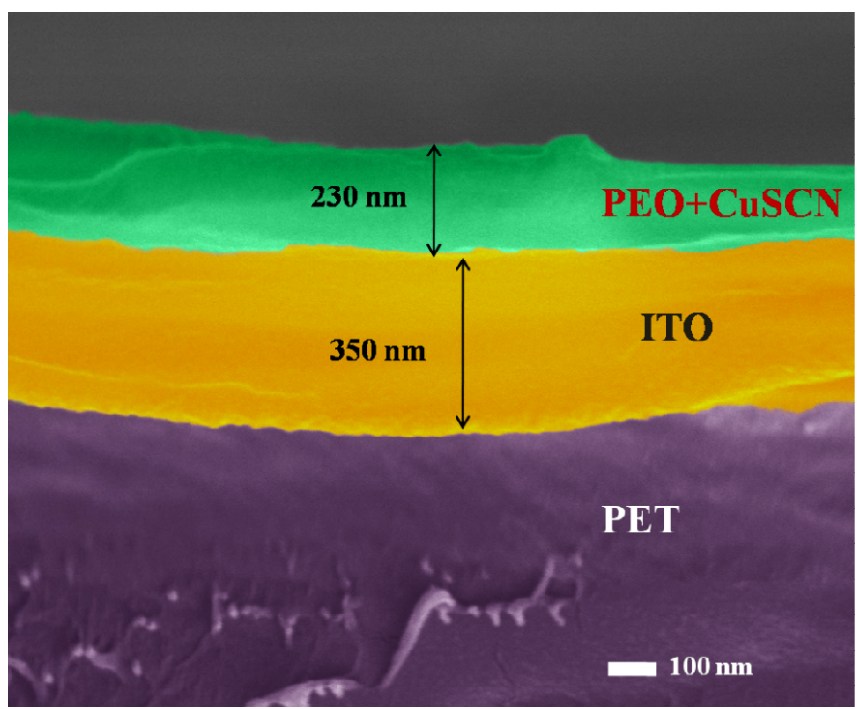

**Figure S5.** Cross-sectional SEM image of Cu-SPE layer deposited over ITO-coated PET substrate at 3000 rpm.

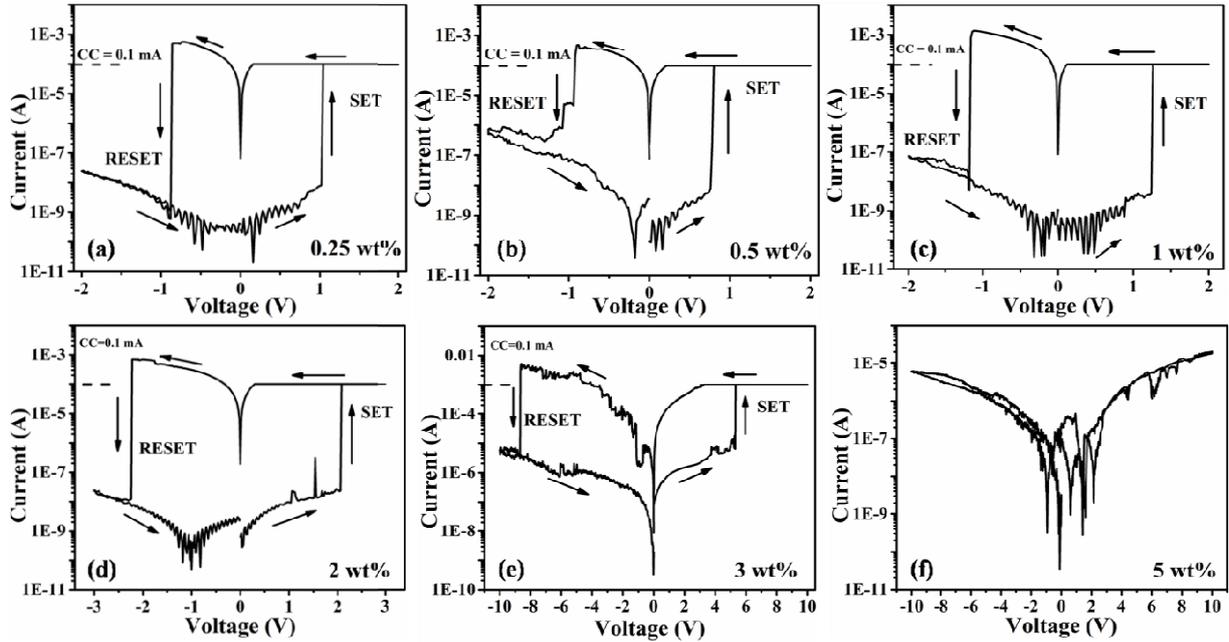

**Figure S6.** *I-V* plots of the ITO/Cu-SPE/Cu memory device with CuSCN concentration of (a) 0.25 wt% (b) 0.5 wt% (c) 1 wt% (d) 2 wt% (e) 3 wt% and (f) 5 wt%.

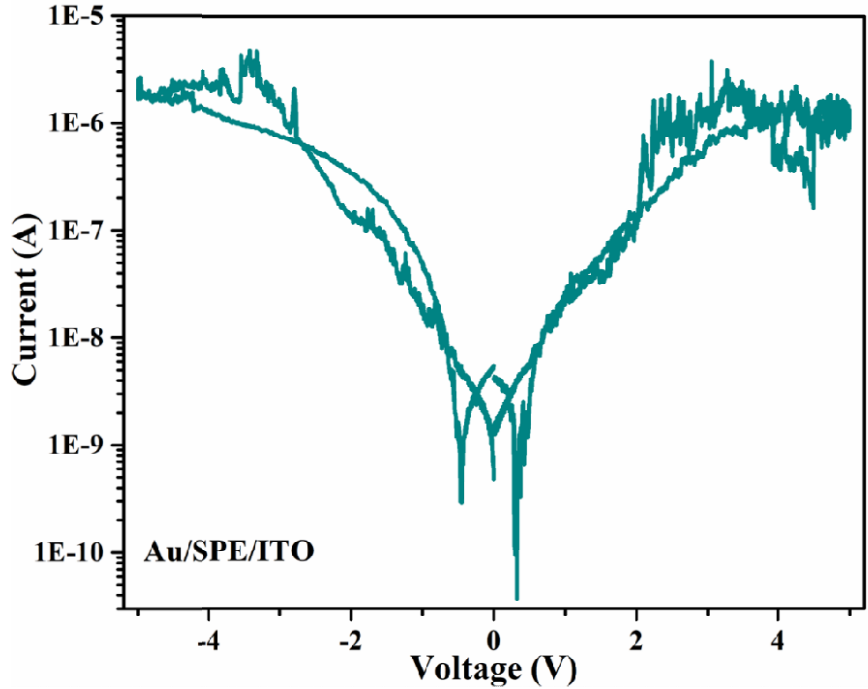

**Figure S7.** Current-voltage (*I-V*) characteristics of Au/SPE/ITO memory device.